\documentclass[amsmath,amssymb,
showpacs,twocolumn,
superscriptaddress,
prl]{revtex4-1}
\usepackage{graphicx,bm,color,subfigure}
\usepackage[T1]{fontenc}
\usepackage{mathrsfs}
\usepackage{bm}
\usepackage{amsmath}
\usepackage{amssymb}
\usepackage{graphicx}
\usepackage{esint}
\usepackage{multirow}
\usepackage{float}
\usepackage{array}
\usepackage{makecell}
\usepackage{harpoon}
\usepackage{booktabs}
\usepackage{gensymb}
\usepackage{simplewick}
\usepackage{subfigure}
\usepackage{soul}


\begin{document}
\title{The s$^\pm$-Wave Pairing and the Destructive Role of Apical-Oxygen Deficiencies in La$_3$Ni$_2$O$_7$ Under Pressure}
\author{Yu-Bo Liu}
\affiliation{School of Physics, Beijing Institute of Technology, Beijing 100081, China}
\author{Jia-Wei Mei}
\email{meijw@sustech.edu.cn}
\affiliation{Shenzhen Key Laboratory of Advanced Quantum Functional Materials and Devices, Southern University of Science and Technology, Shenzhen 518055, China}
\affiliation{Institute for Quantum Science and Engineering and Department of Physics, Southern University of Science and Technology, Shenzhen 518055, China}
\author{Fei Ye}
\affiliation{Shenzhen Key Laboratory of Advanced Quantum Functional Materials and Devices, Southern University of Science and Technology, Shenzhen 518055, China}
\affiliation{Institute for Quantum Science and Engineering and Department of Physics, Southern University of Science and Technology, Shenzhen 518055, China}
\author{Wei-Qiang Chen}
\email{chenwq@sustech.edu.cn}
\affiliation{Shenzhen Key Laboratory of Advanced Quantum Functional Materials and Devices, Southern University of Science and Technology, Shenzhen 518055, China}
\affiliation{Institute for Quantum Science and Engineering and Department of Physics, Southern University of Science and Technology, Shenzhen 518055, China}
\author{Fan Yang}
\email{yangfan_blg@bit.edu.cn}
\affiliation{School of Physics, Beijing Institute of Technology, Beijing 100081, China}
\date{\today}
\begin{abstract}
Recently, the bilayer perovskite nickelate La$_3$Ni$_2$O$_7$ has been reported to show evidence of high-temperature superconductivity (SC) under a moderate pressure of about 14 GPa. To investigate the superconducting mechanism, pairing symmetry, and the role of apical-oxygen deficiencies in this material, we perform a random-phase-approximation based study on a   bilayer model consisting of the $d_{x^2-y^2}$ and $d_{3z^2-r^2}$ orbitals of Ni atoms in both the pristine crystal and the crystal with apical-oxygen deficiencies.  Our analysis reveals an $s^{\pm}$-wave pairing symmetry driven by spin fluctuations. The crucial role of pressure lies in that it induces the emergence of the $\gamma$-pocket, which is involved in the strongest Fermi-surface nesting. We further found the emergence of local moments in the vicinity of apical-oxygen deficiencies, which significantly suppresses the $T_c$.  Therefore, it is possible to significantly enhance the $T_c$ by eliminating oxygen deficiencies during the synthesis of the samples.
\end{abstract}

\maketitle
{\bf Introduction:}
The recent discovery of evidence showcasing superconductivity (SC) with a critical temperature $T_c \approx 80$ K in the Ruddlesden-Popper bilayer perovskite nickelate La$_3$Ni$_2$O$_7$ under a moderate pressure of 14 - 43.5 GPa \cite{HSun} has captured considerable research attention \cite{Haihu2023,Jingguang2023,Yuanhuiqiu2023, ZHLuo, YZhang, QGYang, FLechermann, HSakakibara, YShen, YHGu, VChristiansson, Shilenko2023,XChen2023, Congjun2023, Weili2023,WuWei2023,YangYiFeng2023,ZhangYaHui2023,YangZhangZhang2023}. If verified, these findings could introduce a new family of superconductors with $T_c$ surpassing the boiling point of liquid nitrogen under moderate pressure, distinct from the cuprates \cite{bednorz, anderson, P_A_Lee, keimer, sakaibara, WLi}.

Similar to previously synthesized infinite-layer nickelates \cite{DLi, DLi1, SZeng, MOsada, MOsada1, GAPan, SZeng1}, La$_3$Ni$_2$O$_7$ exhibits a layered structure, wherein the conducting NiO$_2$ layer shares structural similarities with the CuO$_2$ layer found in the cuprates. In the crystal structure of La$_3$Ni$_2$O$_7$, each unit cell comprises two NiO$_2$ layers interconnected by the Ni-O-Ni $\sigma$ bond. This bonding occurs through the participation of an intermediate oxygen atom originating from the intercalating LaO layer. Note that although La$_3$Ni$_2$O$_7$ is an established material\cite{Tsato,DKSeo,YKoba,MGreen1,MGreen2,CDLing,GWu,TFuka,VIVoronin,DOBan,THosoya,bareU,MNakata,YMoch,ZLi,JSong,ZLiu}, its properties are fundamentally changed under pressure as it has induced a structural transition leading to a change in the bonding angle along the $c$-axis from 168\degree to 180\degree \cite{HSun}. This transition also prompts a shift in the crystal's orthorhombic space group from Amam to Fmmm. First-principles calculations indicate that this structural transformation results in the metallization of the $\sigma$ bonding and a concurrent alteration of the band structure from weakly insulating to metallic behavior \cite{HSun}.

In the experiment conducted by Sun et al. \cite{HSun}, it is noteworthy that not all samples of nickelates exhibited SC or even a metallic behavior under pressure. The resistivity is very sensitive to the oxygen content. Certain samples remained insulating and could not transition to a superconducting state. This behavior could be attributed to the presence of slightly oxygen-deficient compositions, specifically in La$_3$Ni$_2$O$_{7-\delta}$. Previous experimental investigations \cite{Goodenough, Taniguchi} have demonstrated that when the oxygen deficiency, denoted as $\delta$, exceeds 0.08, the material transforms into a weak insulator even under ambient pressure. These experimental observations underscore the importance of apical-oxygen deficiency as a non-negligible factor, even in superconducting samples.

In this paper, we aim at providing answers for the following three urgent questions about the pairing nature of the pressurized La$_3$Ni$_2$O$_7$. (1)What is the pairing mechanism and resultant pairing symmetry? (2)What is the role of pressure in realizing the high-$T_c$ SC? (3)How do the apical-oxygen deficiencies affect the $T_c$? To initiate this inquiry, we adopt a bilayer tight-binding (TB) model involving solely $d_{x^2-y^2}$ and $d_{3z^2-r^2}$ orbitals of nickel atoms \cite{ZHLuo}, derived from density-functional-theory (DFT) calculations. This model is supplemented with a standard multi-orbital Hubbard interaction. We first conduct a conventional random-phase-approximation (RPA) calculation in $\mathbf{k}$-space for the pristine crystal. This analysis reveals an s$^{\pm}$-wave pairing symmetry driven by spin fluctuations. The crucial role of pressure lies in that it induces the  $\gamma$-pocket, as the nesting between this pocket and the existing $\beta$ pocket is the strongest. Subsequently, we explore the impact of apical-oxygen deficiencies using a real-space RPA calculation. Our results suggest that in the vicinity of the apical-oxygen deficiencies, due to reduction in the strength of the magnetic critical interaction, magnetic local moments emerge, which strongly suppress the $T_c$. This observation underscores the necessity of eliminating apical-oxygen deficiencies to attain high-T$_c$ SC in pressurized La$_3$Ni$_2$O$_7$.

\begin{figure}[t]
	\centering
	\includegraphics[width=0.5\textwidth]{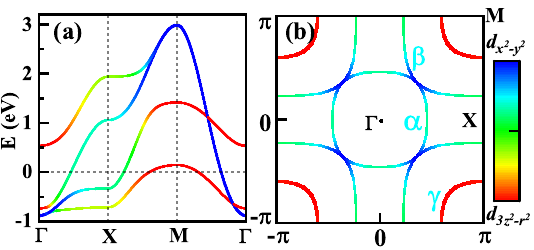}
	\caption{(Color online) (a) Band structure along the high symmetry lines. (b) FS in the first Brillouin zone. The three pockets are labeled by $\alpha, \beta$ and $\gamma$, respectively. The colors in (a) and (b) indicate the orbital component. }\label{band_fs}
\end{figure}

{\bf Model and Approach:} Our two-orbital TB model on a bilayer square lattice reads,
\begin{eqnarray}\label{H_0}
H_0=\sum_{i\mu,j\nu,\sigma}t_{ij}^{\mu\nu}c^{\dagger}_{i\mu\sigma}c_{j\nu\sigma}.
\end{eqnarray}
Here, $i/j$ labels sites, $\sigma$ labels spin, and $\mu/\nu$ labels orbital ($d_{x^2-y^2}$ or $d_{3z^2-r^2}$). The hopping integral $t_{ij}^{\mu\nu}$ is obtained through DFT calculations~\cite{ZHLuo}. The obtained band structure is depicted in Fig.~\ref{band_fs}(a) wherein the Fermi level intersects only
three out of the four bands, giving rise to three Fermi pockets as
illustrated in Fig.~\ref{band_fs}(b). Among these pockets, the one
labeled as $\alpha$ represents an electron pocket, whereas the $\beta$
and $\gamma$ pockets are hole pockets. Notably, the $\alpha$ and
$\beta$ pockets exhibit close proximity to each other. These two pockets display significant hybridization between the two orbitals, whereas the $\gamma$ pocket predominantly
comprises the $d_{3z^2-r^2}$ orbital.  Importantly, DFT calculations
reveal that the $\gamma$ pocket emerges only under pressure,
coinciding with the existence of SC under
pressure. This suggests that the $\gamma$ pocket may play a crucial
role in the SC of the system.

We adopt the following multi-orbital Hubbard interaction,
\begin{eqnarray}\label{H_inter}
H_I &=& U\sum_{i\mu}n_{i\mu\uparrow}n_{i\mu\downarrow}+V\sum_{i\sigma}n_{i1\sigma}n_{i2\sigma}+J_H[\sum_{i\sigma\sigma'}\nonumber\\
& c^{\dagger}_{i1\sigma}& c^{\dagger}_{i2\sigma'}c_{i1\sigma'}c_{i2\sigma}+\sum_{i}c^{\dagger}_{i1\uparrow}c^{\dagger}_{i1\downarrow}c_{i2\downarrow}c_{i2\uparrow}+h.c.].
\end{eqnarray}
Here, the
$U(V)$ term represents the intra (inter)-orbital Hubbard repulsion, and the
$J_{H}$-term represents the Hund’s rule coupling and pair hopping. The Kanamori relation\cite{CCastellani} is adopted, which sets
$U=V+2J_H$, and for the subsequent calculations, we fix
$J_H=U/6$.

\begin{figure}[b]
	\centering
	\includegraphics[width=0.5\textwidth]{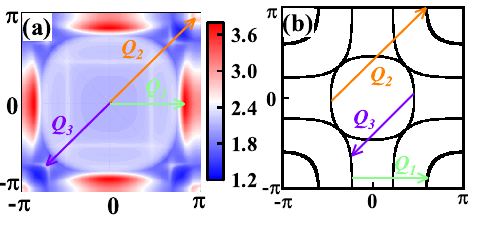}
	\caption{(Color online) (a) The distribution of the largest eigenvalue $\chi(\mathbf q)$ of the RPA-renormalized spin susceptibility matrix in the Brillouin zone. This distribution peaks at three unequivalent momenta, which are marked as $\mathbf Q_1$, $\mathbf Q_2$, and $\mathbf Q_3$ respectively. (b) The FS- nesting with the nesting vectors $\mathbf Q_1$, $\mathbf Q_2$, and $\mathbf Q_3$ marked.}\label{chi0}
\end{figure}

Our study employs the multi-orbital RPA method, utilizing both the
traditional \textbf{k}-space formalism for pristine
crystals\cite{TTakimoto,KYada,KKubo,KKuroki_2008,SGraser,FLiu,XWu_2014,TMa}
and the newly-developed real-space formalism\cite{Ye_Cao} for systems
with randomly distributed apical-oxygen deficiencies. In both cases,
we need first to calculate the bare susceptibilities, then proceed to
determine the renormalized spin and charge susceptibilities at the RPA
level. There exist critical interaction strengths for both charge and
spin, denoted by $U_c^{(c)}$ and $U_c^{(s)}$, respectively.
If $U$ exceeds these critical values, the charge or spin
susceptibility diverges accordingly, and further RPA treatment becomes
invalid. Subsequently, a charge-density wave (CDW) or spin-density
wave (SDW) will
develop.  
In general, $U_c^{(s)}<U_c^{(c)}$. Therefore, when we increase the
repulsive Hubbard $U$, it is more likely for an SDW to form first.
When $U$ is smaller than $U_c^{(s)}$, the short-range spin fluctuation
may act as a ``glue'' that facilitates the formation of Cooper
pairs. This effective attraction can be approximated using the
mean-field (MF) treatment, and by considering this, we can derive a
linearized gap equation in the vicinity of $T_c$.  By discretizing this gap equation, we can transform it into an eigenvalue
problem associated with the interaction matrix.
The largest eigenvalue, denoted as $\lambda$, determines the critical temperature $T_c$  via $T_c\sim e^{-\lambda^{-1}}$. Additionally, the corresponding eigenvector determines the pairing gap function and the pairing symmetry.

{\bf The s$^\pm$-wave pairing:} The critical interaction strength,
denoted as $U_c^{(s)}$, at which the spin susceptibility diverges, is
approximately 1.26 eV following the RPA calculation for the model in
consideration.  We choose the value of the Hubbard interaction $U$ to
be 1.16 eV, in order to match the experimentally measured critical
temperature $T_c$ of approximately 80 K. Note that for the simplified two-orbital model adopted here, this $U$ can be different from the bare $U$ employed in the DFT calculations\cite{bareU}. Fig.~\ref{chi0}(a) depicts
the distribution of the RPA-renormalized spin susceptibility,
represented by $\chi(\mathbf q)$ the largest eigenvalue of the spin
susceptibility matrix, across the Brillouin zone.  Notably, the
distribution exhibits peaks at three unequivalent momenta, which we have
labeled as $\mathbf Q_1$, $\mathbf Q_2$, and $\mathbf Q_3$. These
momenta precisely correspond to the three Fermi surface (FS) nesting
vectors, as illustrated in Fig.~\ref{chi0}(b), and the vector
$\mathbf Q_1$ is associated with the highest intensity of spin
susceptibility.

\begin{figure}[htbp]
	\centering
	\includegraphics[width=0.5\textwidth]{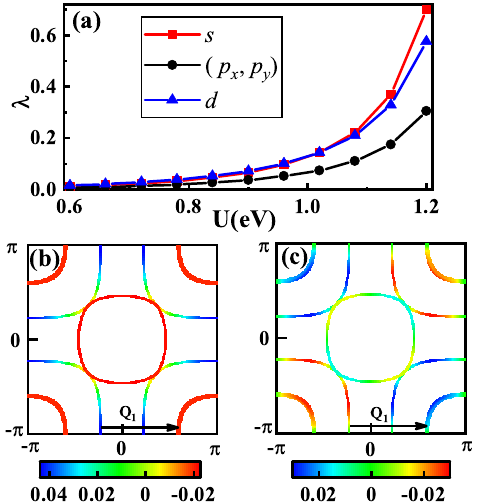}
	\caption{(Color online) (a) The largest pairing eigenvalue $\lambda$ of the various pairing symmetries as function of the interaction strength $U$ with fixed $J_H=U/6$. (b-c) The distributions of the leading $s$ and $d$-wave pairing gap functions on the FS for $U=1.16$eV. }\label{lambda}
\end{figure}

Fig.~\ref{lambda} (a) illustrates the dependence of the largest
pairing eigenvalue, denoted as $\lambda$, on the interaction strength
$U$ for different potential pairing symmetries. The $D_{4h}$ point
group of La$_3$Ni$_2$O$_7$ allows for several possible pairing
symmetries, including non-degenerate $s$-wave, $d$-wave, $g$-wave, and
degenerate $(p_x,p_y)$-wave pairings. However, for clarity, we focus
on the three dominant pairing symmetries in this analysis.  It is
evident that for $U<1$~eV, the $s$-wave and $d$-wave pairings exhibit
similar magnitudes, suggesting a competition between these two
symmetries. On the other hand, for $U>1$eV, the $s$-wave pairing
becomes dominant. Notably, the $(p_x,p_y)$-wave pairing is
consistently suppressed across the entire range of $U$ values
considered.  To further analyze the pairing pattern, we examine the
distribution of the leading $s$-wave gap function on the FS, as depicted in Fig.~\ref{lambda}(b).  Here, we observe that the
gap functions on the $\alpha$ and $\gamma$ pockets of the FS are
negative, while that on the $\beta$ pocket is
positive. This characteristic, known as the $s^\pm$-wave pairing, also verified by the functional renormalization group based studies\cite{QGYang,YHGu}, bears
resemblance to the pairing pattern observed in Fe-based
superconductors\cite{raghu,chubukov,Eschrig,daghofer}.  Additionally,
Fig.~\ref{lambda} (c) shows the distribution of the dominant $d$-wave
pairing gap function on the FS, specifically a $d_{xy}$-wave pairing
symmetry.  These findings provide important insights into the favored
pairing symmetries and shed light on the nature of
\textit{unconventional} superconductivity in La$_3$Ni$_2$O$_7$.

The role of the emergent $\gamma$-pocket under pressure for
SC can be elucidated in three aspects.
Firstly, the Cooper pairing in this system is mediated by spin
fluctuations that peak at momentum $\mathbf Q_1$, as depicted in
Fig.~\ref{chi0}(a). These spin fluctuations arise from the FS nesting, which connects the $\beta$-pocket and the emergent
$\gamma$-pocket, as illustrated in Fig.~\ref{chi0}(b).  Secondly, the
emergence of the $\gamma$ pocket is beneficial for energy gain in the
superconducting state. Fig.~\ref{lambda}(b) and (c) demonstrate that
for both the $s^\pm$- and $d_{xy}$-wave pairings, the gap functions
$\Delta(\mathbf k)$ on FS patches connected by the nesting vector
$\mathbf Q_1$ possess opposite signs. Such a sign distribution of $\Delta(\mathbf k)$ maximizes the energy gain,
similar to the $d_{x^2-y^2}$-wave pairing in the cuprates and the $s^\pm$-wave pairing in the Fe-based superconductors. The $s^\pm$-wave pairing obtained here can be verified by combined phase sensitive experiments\cite{SQUID} and the sharp neutron resonance peak at $\mathbf Q_1$. Thirdly,
the orbital component of the dominant $s^\pm$-wave pairing crucially
relies on the presence of the $\gamma$-pocket. Specifically,
considering the pairing between interlayer (intralayer) $d_{3z^2-r^2}$
orbitals as $\Delta^z_{1}$ ($\Delta^z_{0}$), and the pairing between
$d_{x^2-y^2}$ orbitals as $\Delta^x_{1}$ ($\Delta^x_{0}$), our
findings indicate $\Delta^x_{1}/\Delta^z_{1}=0.59$,
$\Delta^x_{0}/\Delta^z_{1}=-0.18$, and
$\Delta^z_{0}/\Delta^z_{1}=0.17$. These results highlight that the
strongest pairing is between the interlayer $d_{3z^2-r^2}$ orbitals . Importantly, Fig.~\ref{band_fs} (b) shows that the
$d_{3z^2-r^2}$ orbital component is predominantly distributed on the
$\gamma$-pocket.

\begin{figure}[htbp]
	\centering
	\includegraphics[width=0.47\textwidth]{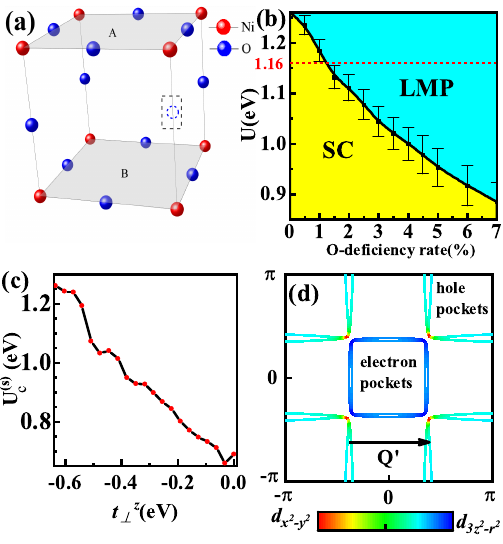}
	\caption{(Color online) (a) Schematic lattice structure with the apical-oxygen deficiency represented by the dotted circle in the black box. (b) The phase diagram with respect to $U$ and the oxygen-deficiency concentration $\delta$. SC(LMP) is short for superconductivity(local moment phase). For each data point, we generate one hundred random deficiency configurations to calculate the statistical expectation value of $U_c^{(s)}$, i.e. $\bar{U}_c^{(s)}$ and mark the error bar. (c) The $t_{\perp}$ dependence of $U_c^{(s)}$ in the uniform system. (d) The FS of the model with $t_{\perp}$ set to zero.}\label{oxygen}
\end{figure}
{\bf Role of apical oxygen-deficiencies:} Previously, it was known that the reduction of the oxygen atoms number within a unit cell from 7 to 6.5 causes the metal-insulator transition\cite{bareU}. Here we examine the impact of a slight concentration of apical-oxygen deficiencies on the system. As shown in Fig.~\ref{oxygen}(a), the oxygen atom between the two layers of Ni-atoms is called apical oxygen, and a certain ratio of apical oxygen may be deficient during sample preparation with insufficient oxygen pressure. In order to simulate the phenomenon, we randomly select some unit cells in which the $t^{z}_{\perp}$ and $t^{x}_{\perp}$ in Eq. (\ref{H_0}) are set as 0, as the apical oxygen deficiency makes it difficult for electrons to hop between two layers. Since we randomly introduce some apical-oxygen deficiencies, the translation symmetry is broken, which necessitates the real-space RPA approach to study the problem. See details of this approach in the Appendix A.

\begin{figure}[htbp]
	\centering
	\includegraphics[width=0.47\textwidth]{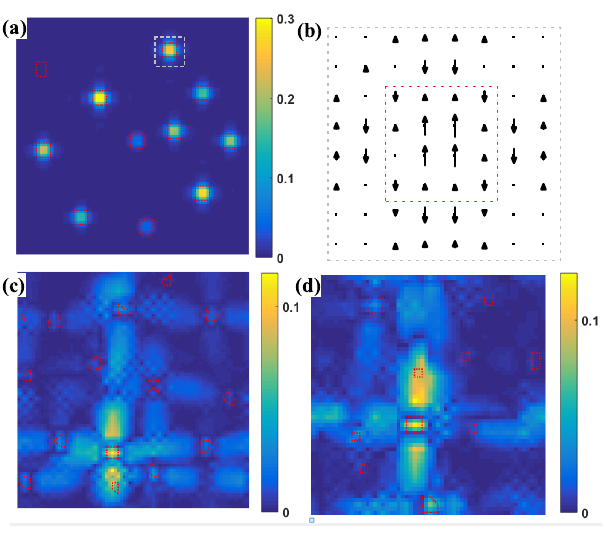}
	\caption{(Color online) (a-c) Real-space distribution of the magnetic moment for the apical-oxygen deficiency concentration $\delta=7\%$, specifically focusing
          on the dominant $d_{3z^2-r^2}$ orbital in the top layer, and that in the bottom layer is oppositely oriented. (a) Unique configuration with segregated regions of high deficiency concentrations.  (b) Zooming in the region
          delineated by white dashed lines in (a) to highlight the magnetic pattern. The arrow indicates the orientation of the magnetic moment. (c) Representative pattern with
          randomly distributed clusters of deficiencies of varying sizes and shapes. Regions with abundant deficiencies are marked by red boxes. In (a) and (c), the color scale represents the magnitude of the magnetic moment.   (d) The $\delta$ dependence of $T_c$.
       } \label{distribution}
\end{figure}

For each configuration of the random distribution of the apical-oxygen deficiencies, we adopt the real-space RPA approach to calculate the magnetic critical interaction strength $U_c^{(s)}$. Then we calculate the statistical expectation value of $U_c^{(s)}$ over a large number of configurations, i.e. $\bar{U}_c^{(s)}$. The deficiency concentration $\delta$ dependence of $\bar{U}_c^{(s)}$ is shown in Fig.~\ref{oxygen}(b) by the black line, in together with its statistical error. Fig.~\ref{oxygen}(b) suggests that $\bar{U}_c^{(s)}$ lowers promptly with the enhancement of $\delta$. When the interaction strength $U$ is below $\bar{U}_c^{(s)}$, the system is superconducting; when $U$ goes beyond $\bar{U}_c^{(s)}$, the system enters into the magnetic phase. For our adopted interaction strength $U=1.16$eV marked by the red dotted line in Fig.~\ref{oxygen}(b), the system enters into the magnetic phase when $\delta\gtrsim 0.02$.

To reveal why the apical-oxygen deficiencies suppress $\bar{U}_c^{(s)}$, we first calculate the $t_{\perp}$ dependence of $U_c^{(s)}$ in the uniform system with reduced $t_{\perp}$, as the apical-oxygen deficiencies reduce $t_{\perp}$ on the average. From the result shown in Fig.~\ref{oxygen}(c), it is clear that $U_c^{(s)}$ is seriously suppressed by the reduction of $t_{\perp}$. The reason for this lies in that the FS-nesting becomes better with reduced $t_{\perp}$, which is clearly shown in the Appendix B. As shown Fig.~\ref{oxygen}(d), the FS for $t_{\perp}=0$ is nearly perfectly nested, and consequently the $U_c^{(s)}$ is reduced almost by half.

However, the suppression of $\bar{U}_c^{(s)}$ by apical-oxygen deficiencies
is not solely attributed to the reduction in averaged $t_{\perp}$. The deficiencies also promote the formation of local moments.  This can be
observed in Fig.~\ref{distribution}(a) where the apical-oxygen
deficiencies are segregated into distinct regions enclosed within red
dotted boxes.  Consequently, the magnetic moment localizes within or
near these regions. Zooming in on the area delineated by the white dashed line
in Fig.~\ref{distribution}(a), which includes a representative region
with significant deficiencies, our real-space RPA calculation yields that, due to reduction of local $U_c$ in this region, magnetic moments emerge, distributed with the pattern shown in Fig.~\ref{distribution}(b). This pattern
reveals a localized antiferromagnetic (AFM) arrangement that
diminishes away from the center of the deficiency. Such localized AFM can cause reduction of the uniform spin susceptibility below the Neel temperature, which can be correlated with previous measurements of magnetic
susceptibility in La$_3$Ni$_2$O$_{6.92}$ \cite{Goodenough}.  These
measurements revealed a slight decrease in magnetic susceptibility
between 100$\sim$300 K.  In Fig.~\ref{distribution}(c), we
show another typical pattern with randomly distributed clusters of
oxygen deficiencies with varying sizes and shapes. Interference occurs between adjacent clusters, resulting in regions without
deficiencies exhibiting significant magnetic moments.

The randomly distributed magnetic moment shown in Fig.~\ref{distribution}(a-c) can significantly harm SC. The $\delta$ dependence of $T_c$ is estimated as follow. For an arbitrary deficiency configuration, if the obtained $U_c<U$, the magnetic moment emerges for that configuration, under which we set $T_c=0$ as an approximation. If $U_c>U$, we adopt the real-space RPA approach provided in the Appendix A to calculate the $T_c$ for this configuration. For our specifically chosen $U=1.16$ eV, the averaged $T_c$ over a large number of deficiency configurations is shown in  Fig.~\ref{distribution}(d) as function of $\delta$, in together with the associated error bar. Remarkably, the $T_c$ is significantly suppressed by the deficiency, and the SC nearly vanishes when $\delta$ surpasses $3\%$. In comparison with the pressurized hydrides whose SC is driven by electron-phonon interaction\cite{YLi,DDuan}, the oxygen deficiencies here play a more destructive role on the $T_c$ due to the pair-breaking effect of the local moment.

{\bf Conclusion:} We have studied the pairing mechanism and pairing symmetry of La$_3$Ni$_2$O$_7$ under pressure by the RPA approach. We find that the pairing symmetry is $s^\pm$, and the gap sign of the $\alpha$- and $\gamma$- pockets are opposite to that of the $\beta$ pocket. The interlayer $d_{3z^2-r^2}$ orbital pairing dominates the pairing. The presence of the emergent $\gamma$-pocket under pressure plays a crucial role in the pairing mechanism as it is involved in the strongest FS nesting.

In addition, we have also studied the effect of apical-oxygen deficiencies through the real-space RPA approach. We find that local antiferromagnetic moment would be generated in the region with rich apical-oxygen deficiencies, which would severely suppress the SC. This conclusion indicates that sufficient oxygen pressure is required to enhance the $T_c$ of the sample.

\appendix

\section{The real-space RPA approach}
Since the system lacks translation symmetry, we should perform RPA approach in the real space. This is different from the usual derivation in the momentum space. So we exhibit the details of the real-space multi-orbital RPA approach as follow.

Let's define the bare susceptibility for the non-interacting case as,
\begin{eqnarray}\label{chi_00}
\chi^{(0)ipq}_{jst}(\tau)\equiv< T_{\tau}c^{\dagger}_{ip}(\tau)c_{iq}(\tau)c^{\dagger}_{js}(0)c_{jt}(0) >_0.
\end{eqnarray}
where $<>_0$ denotes the thermal average for the non-interacting case, $T_\tau$ is the imaginary-time-ordered product. Note that here for the convenience of the derivation, we set $i,j$ only as site indices instead of combined site and layer indices. On the other hand, the $p,q,s,t = 1,\cdots, 4$ are combined orbital-layer indices. These conventions are a little different from those defined in the main text. Fourier transformed to the imaginary frequency space, the bare susceptibility can be expressed as
\begin{eqnarray}\label{chi_0a1}
\chi^{(0)ipq}_{jst}(i\omega)=\sum_{mn}\xi^{*}_{ip,m}\xi_{iq,n}\xi^{*}_{js,n}\xi_{jt,m}\dfrac{n_F(\tilde{\epsilon_m})-n_F(\tilde{\epsilon_n})}{i\omega+\tilde{\epsilon_n}-\tilde{\epsilon_m}}.\nonumber\\
\end{eqnarray}
Here $m,n$ are eigenstate indices and $n_F$ is the Fermi-Dirac distribution function. The $\epsilon$ and $\xi$ are the eigenvalue and eigenstate of the TB model, and $\tilde{\epsilon}=\epsilon-\mu_c$, where $\mu_c$ is the chemical potential.

Considering the interaction in Eq. (\ref{H_inter}), we further define the following spin(s) and charge(c) susceptibilities,
\begin{eqnarray}\label{chi_sc1}
\chi^{(c)ipq}_{jst}(\tau)&\equiv&\frac{1}{2}\sum_{\sigma,\sigma'}< T_{\tau}c^{\dagger}_{ip\sigma}(\tau)c_{iq\sigma}(\tau)c^{\dagger}_{js\sigma'}(0)c_{jt\sigma'}(0) >.\nonumber\\
\chi^{(s)ipq}_{jst}(\tau)&\equiv&\frac{1}{2}\sum_{\sigma,\sigma'}< T_{\tau}c^{\dagger}_{ip\sigma}(\tau)c_{iq\sigma}(\tau)c^{\dagger}_{js\sigma'}(0)c_{jt\sigma'}(0) >\sigma\sigma'.\nonumber\\
\end{eqnarray}
Here $\sigma$ and $\sigma'$ are spin indices. At the RPA level, the renormalized spin and charge susceptibilities read
\begin{eqnarray}\label{chi_sc2}
\chi^{(c/s)}=[I\pm\chi^{(0)}U^{(c/s)}]^{-1}\chi^{(0)},
\end{eqnarray}
where the elements of the matrix $U^{(s/c)}$ are functions of the interaction coefficients, which are as follow,
$$ U^{(s)ipq}_{jst}=\left\{
\begin{aligned}
U , ~~~p=q=s=t \\
J_H,~~~ p=q\neq s=t \\
J_H,~~~ p=s\neq q=t \\
V,~~~ p=t\neq s=q\\
\end{aligned}
\right.
$$

$$ U^{(c)ipq}_{jst}=\left\{
\begin{aligned}
U , ~~~p=q=s=t \\
2V-J_H,~~~ p=q\neq s=t \\
J_H,~~~ p=s\neq q=t \\
2J_H-V,~~~ p=t\neq s=q\\
\end{aligned}
\right.
$$

Similar to the description in the main text, there exist the critical interaction strength $U_c^{(c/s)}$ for charge (c) or spin (s). When $U>U_c^{(c)}$ ($U>U_c^{(s)}$), the charge (spin) susceptibility diverges, implying the onset of charge (magnetic) order, invalidating further RPA treatment. Usually, under repulsive Hubbard interaction, we have $U_c^{(c)}>U_c^{(s)}$, which favors the magnetic order.

From Kubo's linear-response theory, in order to obtain the real-space distribution of the magnetic moments for $U$ slightly stronger than $U_c^{(s)}$, we need to diagonalize $\chi^{(s)ipp}_{jss}$ as $U$ approaches $U_c^{(s)}$. The eigenstate corresponding to the largest eigenvalue of $\chi^{(s)ipp}_{jss}$ determines the real-space distribution of the magnetic moments. When $U$ is weaker than $U_c^{(s)}$, the short-ranged spin fluctuations will mediate effective attraction between a Cooper pair, leading to SC.

\section{Evolution of the FS with $t_{\perp}$}

\begin{figure*}
	\includegraphics[width=7in]{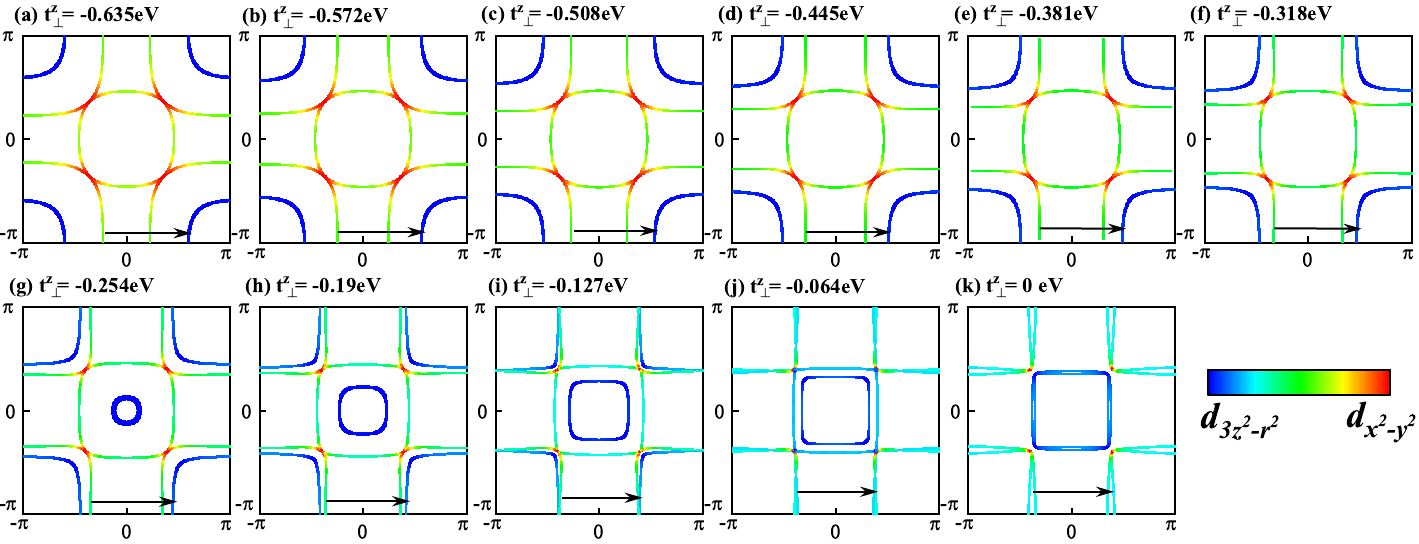}
	\caption{(Color online) The FS evolution with reduced $t_{\perp}$. The FS-nesting vector is marked by the black arrow in each figure. The color indicates orbital component. The value of $t_{\perp}^z$ for each figure has been marked.} \label{appendix1}
\end{figure*}

The FS evolution with reduced $t_{\perp}$ is shown in Fig.~(\ref{appendix1}). The FS-nesting vector is marked in each figure. Obviously, the FS-nesting situation becomes better and better with reduced $t_{\perp}$.

\section*{Acknowledgements}
 This work is supported by the NSFC under the Grant Nos. 12074031, 12234016, 11674025,11774143. W.-Q. Chen and J.W.Mei are supported by the Science, Technology and Innovation Commission of Shenzhen Municipality (No. ZDSYS20190902092905285), Guangdong Basic and Applied Basic Research Foundation under Grant No. 2020B1515120100 and Center for Computational Science and Engineering of Southern University of Science and Technology. J.W.Mei is also supported by the program for Guangdong Introducing Innovative and En- trepreneurial Teams (Grant No. 2017ZT07C062) and Shenzhen Fundamental Research Program (Grant No. JCYJ20220818100405013).


\begin{thebibliography}{10}

\bibitem{HSun}
H. Sun, M. Huo, X. Hu, J. Li, Y. Han, L. Tang, Z. Mao, P. Yang, B. Wang, J. Cheng, D.-X. Yao, G.-M. Zhang, and M. Wang, Nature 10.1038/s41586-023-06408-7 (2023),
arXiv:2305.09586.

\bibitem{Haihu2023} Z. Liu, M. Huo, J. Li, Q. Li, Y. Liu, Y. Dai, X. Zhou, J. Hao, Y. Lu, M. Wang, and H.-H. Wen, arXiv:2307.02950 (2023).

\bibitem{Jingguang2023} J. Hou, P. T. Yang, Z. Y. Liu, J. Y. Li, P. F. Shan, L. Ma, G. Wang, N. N. Wang, H. Z. Guo, J. P. Sun, Y. Uwatoko, M. Wang, G. M. Zhang, B. S. Wang, and J. G. Cheng, arXiv:2307.09865 (2023).

\bibitem{Yuanhuiqiu2023} Y. Zhang, D. Su, Y. Huang, H. Sun, M. Huo, Z. Shan, K. Ye, Z. Yang, R. Li, M. Smidman, M. Wang, L. Jiao, and H. Yuan, arXiv:2307.14819 (2023).

\bibitem{ZHLuo}
Z. H. Luo, X. W. Hu, M. Wang, W. Wu and D. X. Yao, arXiv:2305.15564 (2023).

\bibitem{YZhang}
Y. Zhang, L. F. Lin, Adriana Moreo and Elbio Dagotto, arXiv:2306.03231 (2023).

\bibitem{QGYang}
Q. G. Yang, D. Wang and Q. H. Wang, Phys. Rev. B 108, L140505 (2023).

\bibitem{FLechermann}
F. Lechermann, J. Gondolf, S. Btzel and I. M. Eremin, arXiv:2306.05121 (2023).

\bibitem{HSakakibara}
H. Sakakibara, N. Kitamine, M. Ochi and K. Kuroki, arXiv:2306.06039 (2023).

\bibitem{YHGu}
Y. H. Gu, C. C. Le, Z. S. Yang, X. X. Wu and J. P. Hu, arXiv:2306.07275 (2023).

\bibitem{VChristiansson}
V. Christiansson, F. Petocchi and P. Werner, arXiv:2306.07931 (2023).

\bibitem{Shilenko2023}
D. Shilenko and I. Leonov, arXiv:2306.14841 (2023).

\bibitem{XChen2023}
X. Chen, P. Jiang, J. Li, Z. Zhong, and Y. Lu, arXiv:2307.07154 (2023)

\bibitem{YShen}
Y. Shen, M. P. Qin and G. M. Zhang, arXiv:2306.07837 (2023).

\bibitem{WuWei2023}
W. Wu, Z. Luo, D.-X. Yao, and M. Wang, arXiv:2307.05662 (2023).

\bibitem{YangYiFeng2023}
Y. Cao and Y.-F. Yang, arXiv:2307.06806 (2023).

\bibitem{Congjun2023}
C. Lu, Z. Pan, F. Yang, and C. Wu, arXiv:2307.14965

\bibitem{ZhangYaHui2023}
H. Oh and Y.-H. Zhang, arXiv:2307.15706 (2023).

\bibitem{Weili2023}
X.-Z. Qu, D.-W. Qu, J. Chen, C. Wu, F. Yang, W. Li, and G. Su, arXiv:2307.16873 (2023).

\bibitem{YangZhangZhang2023}
Y.-F. Yang, G.-M. Zhang, and F.-C. Zhang, arXiv:2308.01176 (2023).
	
\bibitem{bednorz}	
J. G. Bednorz and K. A. Muller,  Zeitschrift fur Physik B Condensed Matter 64, 189 (1986).

\bibitem{anderson}
P. W. Anderson, Science 235, 1196(1987).

\bibitem{P_A_Lee}
P. A. Lee, N. Nagaosa, and X.-G. Wen,Rev. Mod. Phys. 78, 17 (2006).

\bibitem{keimer}
B. Keimer, S. A. Kivelson, M. R. Norman, S. Uchida, and J. Zaanen, Nature 518, 179 (2015).

\bibitem{sakaibara}
H. Sakakibara, K. Suzuki, H. Usui, S. Miyao, I. Maruyama, K. Kusakabe, R. Arita, H. Aoki, and K. Kuroki, Phys. Rev. B 89, 224505 (2014).

\bibitem{WLi}
W. Li, J. Zhao, L. Cao, Z. Hu, Q. Huang, X. Wang, Y. Liu, G. Zhao, J. Zhang, Q. Liu, et al., Proceedings of the National Academy of Sciences 116, 12156 (2019).

\bibitem{DLi}
D. Li, K. Lee, B. Y. Wang, M. Osada, S. Crossley, H. R. Lee, Y. Cui, Y. Hikita, and H. Y. Hwang, Nature 572, 624 (2019).
\bibitem{DLi1}
D. Li, B. Y. Wang, K. Lee, S. P. Harvey, M. Osada, B. H. Goodge, L. F. Kourkoutis, and H. Y. Hwang, Phys. Rev.Lett. 125, 027001 (2020).

\bibitem{SZeng}
S. Zeng, C. S. Tang, X. Yin, C. Li, M. Li, Huiong, W. Liu, G. J. Omar, H. Jani, Z. S. Lim, K. Han, D. Wan, P. Yang, S. J. Pennycook, A. T. S. Wee, and A. Ariando, Phys. Rev. Lett. 125, 147003 (2020).

\bibitem{MOsada}
M. Osada, B. Y. Wang, B. H. Goodge, K. Lee, H. Yoon, K. Sakuma, D. Li, M. Miura, L. F. Kourkoutis, and H. Y.Hwang, Nano Letters 20, 5735 (2020).

\bibitem{MOsada1}
M. Osada, B. Y. Wang, B. H. Goodge, S. P. Harvey, K. Lee, D. Li, L. F. Kourkoutis, and H. Y. Hwang, Advanced Materials 33, 2104083 (2021).

\bibitem{GAPan}
G. A. Pan, D. F. Segedin, H. LaBollita, Q. Song, E. M. Nica, B. H. Goodge, A. T. Pierce, S. Doyle, S. Novakov, D. C. Carrizales, A. T. Niaye, P. Shafer, H. Paik, J. T. Heron, J. A. Mason, A. Yacoby, L. F. Kourkoutis, O. Erten, C. M. Brooks, A. S. Botana, and J. A. Mundy, Nature Materials 21, 160 (2021).

\bibitem{SZeng1}
S. Zeng, C. Li, L. E. Chow, Y. Cao, Z. Zhang, C. S. Tang, X. Yin, Z. S. Lim, J. Hu, P. Yang, and A. Ariando, Science Advances 8, eabl9927 (2022).


\bibitem{Tsato} T. Satoshi, N. Takashi, Y. Yukio, K. Yoshiaki, T. Jun, S. Shin-ichi, and S. Masatoshi. J. Phys. Soc. Jpn. 64, 1644 (1995).	

\bibitem{DKSeo} D. K. Seo, W. Liang, M. H. Whangbo, Z. Zhang, and M. Greenblatt. Inorg. Chem. 35 6396 (1996).

\bibitem{YKoba} Y. Kobayashi, S. Taniguchi, M. Kasai, M. Sato, T. Nishioka, and M. Kontani, J. Phys. Soc. Jpn. 65, 3978(1996).

\bibitem{MGreen1} M. Greenblatt. Curr. Opin. Solid State Mater. Sci. 2, 174 (1997).

\bibitem{MGreen2} M. Greenblatt, Z. Zhang, and M. H. Whangbo, Synth. Met. 85, 1451 (1997).

\bibitem{CDLing} C. D. Ling, D. N. Argyriou, G. Wu, and J. J. Neumeier, J. Solid State Chem. 152, 517 (1999)

\bibitem{GWu} G. Wu, J. J. Neumeier, and M. F. Hundley. Phys. Rev. B 63, 245120 (2001).

\bibitem{TFuka} T. Fukamachi, Y. Kobayashi, T. Miyashita, M. Sato. J. Phys. Chem. Solids 62 195 (2001).

\bibitem{VIVoronin} V. I. Voronin, I. F. Berger, V. A. Cherepanov, L. Ya. Gavrilova, A. N. Petrov, A. I. Ancharov, B. P. Tolochko, and S. G. Nikitenko. Nuclear Instruments and Methods in Physics Research Section A: Accelerators, Spectrometers,
Detectors and Associated Equipment 470, 202 (2001).

\bibitem{DOBan} D. O. Bannikov, A. P. Safronov, and V. A. Cherepanov, Thermochim. Acta 451, 22 (2006).

\bibitem{THosoya} T. Hosoya , K. Igawa Y. Takeuchi, K. Yoshida, T. Uryu, H. Hirabayashi and H. Takahashi. J. Phys.: Conf. Ser. 121 052013 (2008).

\bibitem{bareU} V. Pardo and W. E. Pickett, Phys. Rev. B 83, 245128 (2011).

\bibitem{MNakata} M. Nakata, D. Ogura, H. Usui, and K. Kuroki. Phys. Rev. B 95, 214509 (2017).

\bibitem{YMoch} Y. Mochizuki, H. Akamatsu, Y. Kumagai, and F. Oba. Phys. Rev. Mater. 2, 125001 (2018).

\bibitem{ZLi} Z. Li, W. Guo, T. T. Zhang, J. H. Song, T. Y. Gao, Z. B. Gu, and Y. F. Nie. APL Mater. 9, 021118 (2020).

\bibitem{JSong} J. Song, D. Ning, B. Boukamp, J.-M. Bassatd and H. J. M. Bouwmeester. J. Mater. Chem. A, 8, 22206 (2020).

\bibitem{ZLiu} Z. Liu, H. Sun, M. Huo, X. Ma, Y. Ji, E. Yi, L. Li, H. Liu, J. Yu, Z. Zhang, Z. Chen, F. Liang, H. Dong, H. Guo, D. Zhong, B. Shen, S. Li, M. Wang. Sci. China-Phys. Mech. Astron. 66, 217411 (2023).

\bibitem{Goodenough} Z. Zhang, M. Greenblatt, and J. B. Goodenough, J. Solid State Chem. Solids. 108, 402 (1994).
\bibitem{Taniguchi} S. Taniguchi, et al., J. Phys. Soc. Jpn. 64, 1644 (1995).

\bibitem{CCastellani}
C. Castellani, C. R. Natoli, and J. Ranninger, Phys. Rev. B 18, 4945 (1978).




\bibitem{TTakimoto}
T. Takimoto, T. Hotta, and K. Ueda, Phys. Rev. B 69, 104504 (2004).

\bibitem{KYada}
K. Yada and H. Kontani, J. Phys. Soc. Jpn. 74, 2161 (2005).

\bibitem{KKubo}
K. Kubo, Phys. Rev. B 75, 224509(2007).

\bibitem{KKuroki_2008}
K. Kuroki, S. Onari, R. Arita, H. Usui, Y. Tanaka, H. Kontani, and H. Aoki, Phys. Rev. Lett. 101, 087004(2008).

\bibitem{SGraser}
S. Graser, T. A. Maier, P. J. Hirschfeld and D. J. Scalapino, New J. Phys.11, 025016 (2009).

\bibitem{FLiu}
F. Liu, C.-C. Liu, K. Wu, F. Yang and Y. Yao, Phys. Rev. Lett. 111, 066804 (2013).

\bibitem{XWu_2014}
X. Wu, J. Yuan, Y. Liang, H. Fan and J. Hu, Europhys. Lett. 10827006 (2014).

\bibitem{TMa}
T. Ma, F. Yang, H. Yao and H. Lin, Phys. Rev. B 90, 245114 (2014).

\bibitem{Ye_Cao}
Y. Cao, Y. Zhang, Y.-B. Liu, et al, Phys. Rev. Lett. 125, , 017002 (2020).

\bibitem{raghu}
S. Raghu, X.-L. Qi, C.-X. Liu, D. J. Scalapino, and S.-C. Zhang, Phys. Rev. B 77, 220503 (2008).

\bibitem{chubukov}
A. V. Chubukov, D. V. Efremov, and I. Eremin, Phys. Rev. B 78, 134512 (2008).

\bibitem{Eschrig}
H. Eschrig and K. Koepernik, Phys. Rev. B 80, 104503(2009).

\bibitem{daghofer}
M. Daghofer, A. Nicholson, A. Moreo, and E. Dagotto, Phys. Rev. B 81, 014511 (2010).

\bibitem{SQUID} Van Harlingen, Rev. Mod. Phys. 67, 515 (1995).

\bibitem{Supplement} See the Supplementary Materials at ...... for the real-space RPA approach and the evolution of the FS with $t_{\perp}$.


\bibitem{YLi} Y. Li, J. Hao, H. Liu, Y. Li, and Y. M. Ma, J. Chem. Phys. 140,174712 (2014).

\bibitem{DDuan} D. Duan, Y. Liu, F. Tian, D. Li, X. Huang, Z. Zhao, H. Yu, B. Liu, W. Tian, and T. Cui, Sci. Rep. 4, 6968 (2014).


	
\end{thebibliography}
\end{document}